# Enhancing Four-Wave-Mixing Processes by Nanowire Arrays Coupled to a Gold Film

Ekaterina Poutrina,* Cristian Ciraci, Daniel J. Gauthier,[†] and David R. Smith

*Center for Metamaterials and Integrated Plasmonics, Electrical and Computer Engineering and [†]Department of Physics, Duke University, P. O. Box 90291,Durham, NC 27708, USA; ekaterina.poutrina@duke.edu*

We consider the process of four-wave mixing (FWM) in an array of gold nanowires strongly coupled to a gold film. Using full-wave simulations, we perform a quantitative comparison of the FWM efficiency associated with the thin film geometry with and without the nanowire arrays. When the nanowire array is present, both the delocalized surface plasmon of the film as well as the localized surface plasmon resonances of the nanowires contribute to the local field enhancement, yielding an overall FWM efficiency enhancement of up to four orders-of-magnitude over a wide range of excitation angles. The film-coupled nanowire array is easily amenable to nanofabrication, and could find application as an ultra-compact component for integrated photonic and quantum optic systems.

Four wave mixing (FWM) is a nonlinear optical process that has found numerous practical applications, ranging from optical processing [1-3], nonlinear imaging [4, 5], real-time holography and phase-conjugate optics [6-8], phase-sensitive amplification [9,10], and entangled photon pair production [11,12]. Because material nonlinear susceptibilities are relatively small, high conversion efficiency in FWM processes requires that light travels over long interaction lengths—available, for example, in silica fibers [6,10,12]—with considerable effort needed to achieve proper phase-matching conditions. Alternatively, the conversion efficiency can be increased by concentrating optical fields within high-refractive-index materials, such as in silicon waveguides [2,3,11,13]. In the latter case, the spectral range of operation can be affected by two-photon absorption, making the mid-infrared wavelength range the most efficient regime for FWM and other nonlinear processes [13].

Here, we propose a method for increasing efficiency of FWM processes by using metals, which exhibit some of the largest values for $\chi^{(3)}$ as compared with other materials [14,15]. However, it is difficult to exploit the intrinsic nonlinearities of metals because light is mostly reflected from metal surfaces and does not interact significantly with the bulk material. We predict that this problem can be overcome and extremely large third-order susceptibilities realized by coupling free-space light beams into and out of surface plasmons and localized surface plasmon resonances created via nanoscale metallic structures. Both the localized fields and the radiative power of light generated via FWM process is shown to increase by many orders-of-magnitude.

The third-order nonlinear susceptibility in metals arises from both the nonlinear polarization induced by the motion of bulk valence electrons as well as from the surface polarization. The latter contribution can be significantly increased in metals when surface plasmons (SPs) are excited on the metal surface. Under proper excitation conditions, the field strengths associated with SP modes can greatly exceed the field strengths of the incident light, such that even the small interaction volumes associated with thin metal films can substantially increase the overall efficiency of FWM and other nonlinear processes. Additionally, the enhanced fields associated with SPs penetrate into the metal, increasing the volume of the bulk nonlinear interaction. Combined, these effects can produce extremely large enhancements in the effective values of $\chi^{(3)}$. The nanoscopic nature of the interaction regions suggests the thin metal film can operate as an extremely compact device for integrated photonic and quantum optic applications.

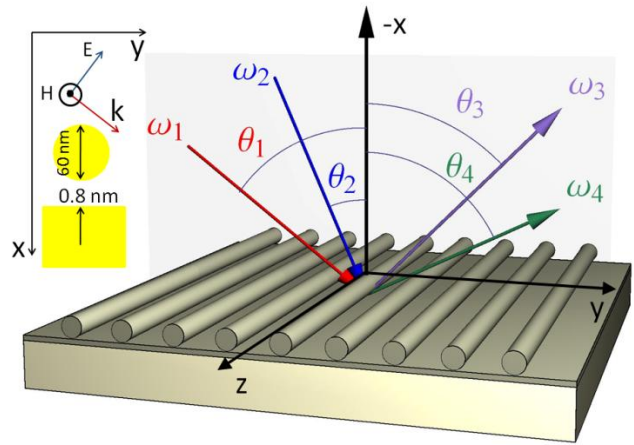

FIG. 1 Schematic of the gold film geometry with array of nanowires. The nanowires are assumed to be spaced 0.8 nm from the gold film surface. Two waves with frequencies $\omega_1$ and $\omega_2$ are incident on the film, producing two mixing components with frequencies $\omega_3$ and $\omega_4$. The inset shows the dimensions of a single nanowire element over the metal film.

Recently, it has been demonstrated that the inherent nonlinearity associated with a gold film can be used to couple light from free space into SPs via a FWM process [16, 17]. SPs are characterized by a wave vector directed along the interface of the film and whose magnitude is larger than the magnitude of the free-space wave vector. In general, light incident onto a planar film from free space cannot excite SPs. However, if the nonlinear polarization contribution along the metal surface is significant, two pump waves incident on the metal surface at frequencies $\omega_1$ and $\omega_2$ can undergo a FWM process, producing a wave at frequency $2\omega_1 - \omega_2$, whose corresponding wave vector can be matched to the that of the SP through a proper choice of incident angles. The planar metallic film is a relatively simple geometry whose SP mode structure is well understood, and which can be readily simulated, fabricated and experimentally investigated.

While leveraging the delocalized SPs of a film provides one path to increase the local field strength, a second path is to introduce nanostructures that support *localized* SP resonances (LSPRs). The local fields associated with LSPRs that occur between interacting metallic nanostructures can provide enhancements of the incident field by several orders-of-magnitude [18-20]. A system of particular relevance here is a collection of metallic nanoelements with nanometer spacing from a metal film surface [21-24]. Extremely large local fields exist between a nanoelement coupled to its electromagnetic image in a metal film, with field enhancements of two or more orders-of-magnitude possible. By arranging the nanoelements into a periodic array on a film surface, it is possible to achieve efficient grating coupling of the incident beams to the delocalized SPs, while simultaneously exciting the LSPRs of the individual elements [25-27]. A specific example of nanowires (NW) coupled to a thin film is illustrated in Fig. 1. In this work, we show that the collective effect of the periodic NW arrangement in such a system, leads to an interplay between the LSPR and SP modes, resulting in many orders-of-magnitude enhancement of both the peak intensity and the radiated power of the generated FWM light.

Motivated by the geometry studied in [17], we assume two *p*-polarized pump waves at frequencies $\omega_1$ and $\omega_2$ incident at angles $\theta_1$ and $\theta_2$ on the film surface, respectively, as shown in Fig. 1. We consider the FWM process involving the frequencies $\omega_3 \equiv 2\omega_1 - \omega_2$ and $\omega_4 \equiv 2\omega_2 - \omega_1$, where the dominant contributions to the nonlinear polarization arise from the terms

$$\mathbf{P}^{NL}(\omega_3) = \varepsilon_0 \chi^{(3)}(-\omega_3; \omega_1, \omega_1, -\omega_2) \vdots \mathbf{E}_1 \mathbf{E}_1 \mathbf{E}_2^*$$
$$\mathbf{P}^{NL}(\omega_4) = \varepsilon_0 \chi^{(3)}(-\omega_4; \omega_2, \omega_2, -\omega_1) \vdots \mathbf{E}_2 \mathbf{E}_2 \mathbf{E}_1^*. \quad (1)$$

Here, $\mathbf{E}_i$ are the electric field vectors of the incident beams located in the *xy* plane and $\chi^{(3)}$ is the third-order nonlinear susceptibility tensor with the triple-dot sign denoting the summation over the relevant field and frequency components. Because we do not aim at analyzing the respective contributions of surface versus volume processes, we treat the total nonlinear response as coming from the bulk with an effective third-order susceptibility of $\chi^{(3)}_{1111} = 0.2 \, \text{m}^2/\text{V}^2$ [18], assuming an isotropic response for the gold medium, such that $\chi^{(3)}_{1111} = (\chi^{(3)}_{1212} + \chi^{(3)}_{1221} + \chi^{(3)}_{1122})$. In the latter expression, the indices 1 and 2 represent any of the Cartesian coordinates *x*, *y*, or *z* as shown in Fig. 1, and we assume $\chi^{(3)}_{1212} = \chi^{(3)}_{1122} = \chi^{(3)}_{1221}$.

As a starting point, we provide the analytical description of FWM process in case of a plain metal film. We subsequently use the analytical results to demonstrate a good accuracy of the employed simulations procedure. From momentum conservation and assuming air for the medium above the surface, the *x*-component of the propagation

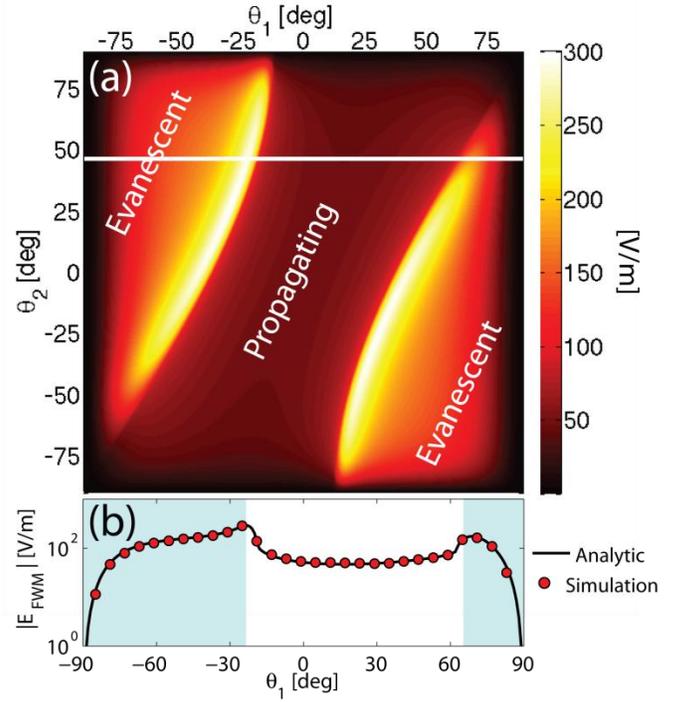

FIG. 2 (a) Intensity of the generated FWM light from a plain gold film as a function of excitation angles. (b) Analytical and numerical results for the norm of the generated FWM field along the white line in (a). An excitation amplitude of 40 MV/m for each of the pump beams, with $\lambda_1 = 612$ nm and $\lambda_2 = 780$ nm is assumed for the calculation.

constant above the film for the generated field at $\omega_3$ can be found as $k_{3,x} = \sqrt{\omega_1^2/c^2 - k_{3,y}^2}$. In the latter expression, accounting for the continuity of the fields along the interface, $k_{3,y} = 2\omega_1/c_1 \sin(\theta_1) - \omega_2/c_1$, with the signs of the angles taken as positive in the direction counterclockwise from the surface normal. In the case of a plain film geometry, $k_{3x}$ has purely imaginary values over a range of pump angles $\theta_1$ and $\theta_2$ (we term this the *evanescent* regime); for these pump wave configurations, the generated FWM light is directed along the surface, with a momentum that coincides with that of a SP. The expressions for the *x*- and *y*-components of the generated FWM field in the region above the metal film can be found as [14]

$$E_x^R = -\frac{k_y^R}{\varepsilon_0 (k_T^2 - k_S^2)(k_x^T - \varepsilon_r k_x^R)} \Big[ k_T^2 P_y^{NL}$$
$$- k_y^S (k_x^S P_x^{NL} + k_y^S P_y^{NL}) - k_x^T (k_x^S P_y^{NL} - k_y^S P_x^{NL}) \Big];$$
$$E_y^R = \frac{k_x^R}{\varepsilon_0 (k_T^2 - k_S^2)(k_x^T - \varepsilon_r k_x^R)} \Big[ k_T^2 P_y^{NL} \quad (2)$$
$$- k_y^S (k_x^S P_x^{NL} + k_y^S P_y^{NL}) - k_x^T (k_x^S P_y^{NL} - k_y^S P_x^{NL}) \Big],$$

where the indices *R* and *T* refer to the to the waves for the generated FWM field, in air and inside the metal respectively, and *S* to the nonlinear source field inside the medium, as shown in the inset in Fig. 1 [14, 17a]. A contour map showing the FWM efficiency at $\omega_3$ calculated according to Eqs. (2) is shown in Fig. 2a. An increase in the intensity of the generated FWM light is seen in the evanescent regimes of the FWM process, consistent with the results described in [17].

To show the dramatic enhancements to $\chi^{(3)}$ that can be attained with the NW-film system that is not amenable to analytical approaches, we make use of full-wave, finite-element based simulations using Comsol Multiphysics. The simulations are performed in two dimensions in the frequency domain, solving four coupled wave equations for the pump fields at $\omega_1$ and $\omega_2$, and for the generated fields at $\omega_3$ and $\omega_4$, coupled through the nonlinear polarization term [28]. We account for self-phase (SPM) and cross-phase modulation (XPM) in the pump equations [28]. For the generated fields at $\omega_3$ and $\omega_4$, we account for the coupling of each of the fields with the pump waves in addition to the nonlinear polarization terms introduced in Eqs. (1); the SPM and XPM effects between the generated fields are neglected.

As a confirmation of the numerical approach, we first simulate FWM from a gold film in the absence of NWs. The thickness of the gold film is taken as 200 nm, much larger than the ~50 nm skin depth of gold over the frequency range considered. Thus, for the purposes of the simulations presented here, the film approximates an infinite half-space such that the bare film should be described by Eqs. 2. The results of a numerical simulation with the angle of the second pump $\theta_2$ fixed at 46°, thus moving along the white line in Fig. 2a, is shown in Fig. 2b and is in excellent agreement with the analytical result in Fig. 2a.

To calculate the FWM efficiency, the maximum of the norm of the electric field is evaluated over the computational domain for each of the excitation angles. The norm of the field is found to be everywhere uniform in the propagating regime, while a maximum is observed in a narrow region along the film surface for pump angle combinations at which SPs are excited (evanescent regime). As seen in Fig. 2b, the efficiency of the FWM process is enhanced around -23° and 69°, the transition regions between the propagating and evanescent regimes, in agreement with Ref. [18]. Both the shape of the angular dependence and the amplitude of the enhancement differ however from those in Refs. [18] and [29], which is a consequence of our use of different pump frequencies and a different angular arrangement. A spectrum similar to that found in Refs. [18] and [29] is produced when their pump configuration. The difference in the angular dependence for the two cases indicates the strong dependence of the FWM efficiency on the dispersion of gold. The particular choice of the wavelength and the angles in calculating Fig. 2 is dictated by the linear scattering spectrum of the NW-film system, as explained below.

We now consider introducing periodically arranged NWs coupled to the metal film in the simulations geometry. The linear scattering spectrum for the cases of NWs spaced 0.8

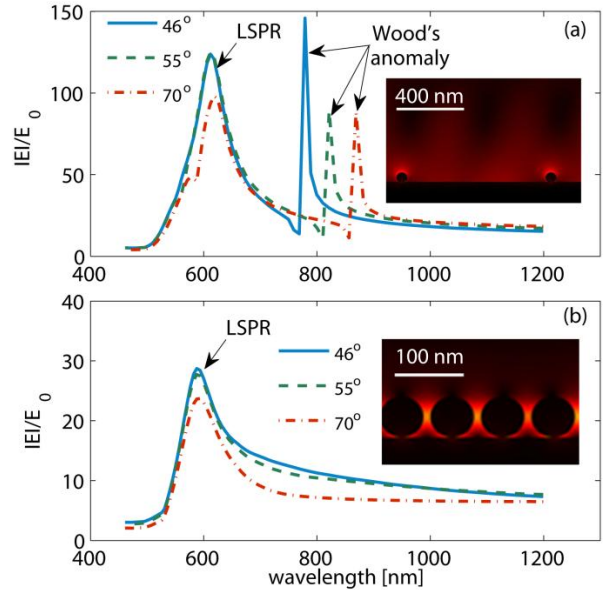

Fig. 3. (a) Linear spectral response of the NW-film system with a 880 nm period spaced 0.8 nm from the film; (b) same as (a) for a 70 nm period NW spacing. The insets show the norm of the electric field for the first Wood's anomaly peak (a) and for the off-resonant excitation with a small period (b)

nm from the film surface with periods of either 880 nm or 70 nm NW is showm in Fig. 3 for several excitation angles. The periodicity values are chosen to allow for the analysis of both the spacing that is comparable and the one that is much smaller than the excitation wavelength. The chioce of the particular numbers is arbitrary in each case and is taken for the convenience of the simulations procedure. With either period, an LSPR resonance is observed around 600 nm, characterized by a strong localized field enhancememnt of roughly two orders-of-magnitude (confined between the NWs and film). The position of this resonance is nearly angle-independent, being defined by the separation of the NW from the film, the material parameters, and the NW diameter (60 nm) [25, 27, 28], and partially by the lattice period: a smaller period leads to larger NW concentration, increasing the effective refractive index above the film, hence blueshifting the resonance [27] in case of the 70 nm periodicity. Additionally, when the lattice constant of the NWs is on the order of a wavelength, Wood's anomalies are produced [30, 31], characterized by narrow, asymmetrically-shaped spectral peaks. Near these anomalies, strong enhancements of the pump field occur, as seen in Fig. 3a, with the peak field similarly localized in the gap.

The spectral positions of the Wood's anomalies are strongly angle-dependent, in contrast with that of the LSPR whose peak remains fixed at 612 nm for all excitation angles. As can be seen from Fig. 3a, the spectral characteristics of the periodic NW-film system are quite complex, and suggest numerous possibilities for the enhancement of nonlinear processes. As a first example, we position the frequency of one of the pump waves at the Wood's anomaly peak, while the frequency of the second pump wave is positioned at the peak of the LSPR resonance (612 nm). Since the Wood's anomaly resonance is angle-dependent, the first pump wave is kept at a fixed angle of $\theta_1 = 46°$ which leads to the first Wood's peak at 780 nm. The angle for the second pump is swept between -90 and 90 degrees, thus again following the white line in Fig. 2a.

The angular dependence of the FWM efficiency at $\omega_3$ ($\lambda_3$=503 nm) is shown as the black curve in Fig. 4a. The presence of a NW array and the optimized pump arrangement that exploits the resonances in the linear response of the NW-film system, leads to an enhancement of the intensity of the produced FWM field by five to nine orders-of-magnitude over the angular range, compared with the bare film. The two minimas located symmetrically at ±26° are due to the equivalent minimas appearing in the linear spectrums of the pump wave placed at the LSPR resonance. With a 26° angle of incidence, the Wood's-anomaly resonance blue-shifts compared with larger angles of incidence and strongly modifies the linear spectral shape of the SP resonance The spectral dips can be eliminated by using other pump configurations, leading to a uniform, broad-angle enhancement of several orders-of-magnitude for the generated light. An alternative configuration to shift the minima is shown as the dashed-red curve in Fig 3c. For this case, the NWs are densely arranged (period 70 nm) on the film surface. The Wood's-anomaly peaks in the linear pump wave spectrum are absent for the dense NW array, thus producing a flat spectrum as a function of angle. To investigate the possibility of FWM enhancement in this configuration, both pump waves are positioned within the LSPR spectrum, at 588 nm and 600 nm, producing a FWM wave at 577 nm with the same angle arrangement as used before.

The overall linear enhancement is smaller for the dense NW array, as seen in Fig. 3b, leading to a somewhat reduced efficiency of the FWM process as compared with the sparse NW array. The overall angle dependence of the FWM process is very uniform in this case, also showing about a six to eight orders-of-magnitude increase compared to the plain film. The greater peak enhancements of the linear fundamental field observed for the sparse array can be attributed to the presence of a standing wave in the linear SP spectrum that occurs for certain wavelengths relative to the array periodicity. An example of such a standing-wave that appears as a linear response of the NW-film system when excited at the first Wood's anomaly peak at 46° (780 nm), is shown in the inset in Fig. 3a. The coupling to this standing wave mode decreases to zero as the excitation angle approaches zero, however, because the component of the field normal to the surface vanishes [23] and the NW-film

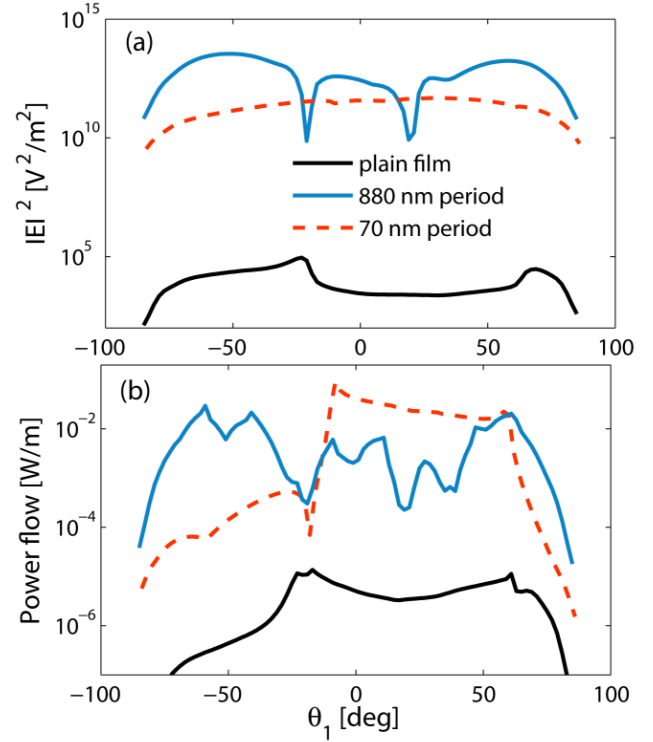

Fig. 4. Enhancement of the FWM efficiency in a periodic arrangement of NWs on a metal surface with a 0.8 nm spacing. (a) FWM peak intensity for the 880 nm (*solid blue*), 70 nm (*dashed red*), and plain film (*solid black*) systems; (b) Power flow for the same cases as in 4a.

LSPR cannot be excited. By contrast, the broadband propagating mode formed between NWs in the dense array can be excited by waves incident from any angle, leading to a FWM efficiency that is much more uniform over excitation angles. An example of such a mode for the fundamental field in the off-resonant regime (at 800 nm) is shown in the inset in Fig. 3b.

While a strong enhancement of the FWM process for the NW-film arrangement, compared to a bare film, is observed in Fig. 4a, the enhanced field is highly localized to the nanoscale regions between the NWs and the film. It is of a practical interest to estimate the available radiated power of the produced FWM field in each configuration. To estimate the power available for extraction, we calculate the power flow of the generated FWM field across the boundaries of the simulation domain. Equal domains with the size of 1,760 nm along the film and 950 nm normal to the film are used in all cases, with the integration performed over the boundary parallel to the film and one of the boundaries normal to the film (the second normal to the film boundary provides the same power flow with an opposite sign due to the periodicity).

The available power flow of the generated FWM field is shown in Fig. 4b for the NW-film arrays with 880 nm (solid blue curve) and 70 nm (dashed red) periods. We note that, in the case of the NW-film array, the propagating plasmon at frequency $\omega_3$ can be excited in either the forward or backward direction along the film, depending on the excitation angle and the NWs period. The grating-like effect of the periodic NW arrangement adds an additional momentum along the metal surface. Thus, to estimate the total radiated power, we integrate the absolute values of the power flow over each of the boundaries, independent of the direction along the film. This leads to the additional structure observed in the power flow curve for the 880 nm period system (solid blue cure in Fig. 4a), compared with the field intensity spectrum of Fig. 4a. As seen from the figure, the radiated power is enhanced by several orders of magnitude in all cases. In addition, the dense NW array provides a similar angular distribution of the FWM efficiency as that for the bare film, but is enhanced by two to four orders-of-magnitude.

In conclusion, we have shown that it is possible to enhance - by many orders of magnitude - FWM process using an array of nanowires strongly coupled to a metal film. Both the increase of up to nine orders of magnitude in the localized fields and of up to four orders of magnitude in the radiative power of the generated light is observed. The strongest enhancement is observed when the local enhancement associated with an LSPR intermingles with the enhancement from the delocalized SP; both modes can interact by arranging the NWs periodically on the surface of a metal film. Given the numerous resonances occurring in the spectrum of the linear response of the metal to the applied fields, tuning of the angles of the incident pump beams can lead to even greater enhanced nonlinear processes, especially where the SP or Wood's anomalies overlap with the LSRP of the NWs, opening a further path to designing a NL media with the enhanced properties .

We thank Jack Mock for valuable discussions. This work was supported by the Air Force Office of Scientific Research (Grant No. FA9550-09-1-0562).